\documentclass[a4paper,preprintnumbers,showpacs,twocolumn,superscriptaddress,nofootinbib,amsmath,amssymb,floatfix]{revtex4-1}

\usepackage{amsmath,amssymb,graphicx,amsfonts}
\usepackage{mathrsfs}
\usepackage{comment}
\usepackage{xcolor}
\usepackage[shortlabels]{enumitem}
\usepackage{url}
\usepackage{graphicx}
\usepackage{dcolumn}
\usepackage{bm}
\usepackage{wasysym}
\usepackage{color}
\usepackage{multirow}
\usepackage{booktabs}
\usepackage{float}

\newcommand{\Pb}{$^{208}$Pb}
\newcommand{\Ca}{$^{40}$Ca}

\begin{document}

\title{\textbf{Coulomb and nuclear polarization of the nucleon density in the dinuclear configuration of the $^{40}$Ca + $^{208}$Pb system}}

\author{Bakhodir Kayumov}
\email{b.kayumov@newuu.uz}
\affiliation{Department of Physics, New Uzbekistan University, Tashkent 100000, Uzbekistan}
\affiliation{Laboratory of Theoretical Nuclear Physics, Institute of Nuclear Physics, Tashkent 100214, Uzbekistan}

\begin{abstract}
The change of the nucleon density distribution of a heavy nucleus caused by the presence of a second nucleus at a fixed centre-to-centre distance is studied in the framework of the deformed Woods--Saxon mean field. The single-particle problem of one nucleus has been solved by diagonalization in a cylindrical harmonic oscillator basis with the external field of the partner nucleus included in the Hamiltonian before diagonalization. The external field consists of the Coulomb potential of the partner charge distribution and of the nuclear part obtained by double folding of the Migdaldensity-dependent effective nucleon-nucleon interaction, in the same form as it is used in the dinuclear system approach. The induced density change $\delta\rho(\bm r;R)$ and the corresponding changes of the quadrupole moment, of the mean-square radius and of the interaction energy have been calculated for the \Ca$+$\Pb~system at separations $R = 15.5$--$20$ fm. The results are compared with the first-order (linear-response) treatment of the same external field. The polarization of both partners is found to be dominated by the repulsive Coulomb tidal field, which produces an oblate deformation of the density, while the attractive nuclear field acts in the opposite sense and partially cancels the Coulomb contribution at the smallest separations. The polarization energy is found to be a dipole quantity, the induced quadrupole deformation contributing only a few percent of it. The limits of validity of the one-centre description are established quantitatively.
\end{abstract}

\maketitle

\section{Introduction}

The synthesis of superheavy elements and the study of the fusion-fission dynamics of massive nuclear systems remain among the active topics of low-energy nuclear physics \cite{Hofman2000,Giuliani2019}. In the entrance channel of such reactions the two nuclei approach each other under the combined action of the Coulomb repulsion and of the nuclear attraction, and at distances close to the touching configuration they form a dinuclear system (DNS) which evolves in mass and charge asymmetry \cite{Antonenko1995,Adamian2000}. The theoretical description of this stage requires the nucleus-nucleus interaction potential, the driving potential and the single-particle level scheme of the two-centre system \cite{Kayumov2022,Nasirov2024,Kayumov2025}. In the DNS approach all of these quantities are constructed from the ground-state densities of the two nuclei, which are kept frozen during the approach, and the deformation parameters entering them are the static ones taken from the tables \cite{AdamianDNS1996,Nasirov2005}. The polarization of the densities by the field of the partner is not taken into account in the equations of relative motion, and the size of the correction which is thereby neglected has not been established microscopically.

This assumption is convenient but it is not exact. Each nucleus is a polarizable object, and the field created by the partner modifies its single-particle wave functions and therefore its density distribution. It is well known that the polarization of the nuclear surface influences the height and position of the Coulomb barrier and the value of the nucleus-nucleus potential in the region of the potential pocket \cite{Denisov2002}.

The frozen-density approximation has been examined from several directions. In the double-folding model the interaction potential is obtained by folding the ground-state densities of the two nuclei with an effective nucleon-nucleon interaction \cite{Satchler1979}, and the generalized version of the model with a density-dependent M3Y force reproduces elastic and inelastic nucleus-nucleus scattering over a wide range of systems \cite{Khoa2000}. The same construction, with the Migdal density-dependent interaction, provides the nucleus-nucleus potential of the DNS approach \cite{Adamian2000,AdamianDNS1996}. In all of these treatments the densities entering the folding integral are those of the isolated nuclei. The frozen Hartree-Fock method makes the same assumption within a self-consistent mean field, the two colliding nuclei keeping their ground-state densities at every separation, and it has been used extensively to construct bare interaction potentials and to study the role of the Pauli principle in the barrier region.

Fully dynamical microscopic descriptions relax this assumption. The time-dependent Hartree-Fock (TDHF) theory follows the self-consistent evolution of the single-particle wave functions of the whole system and therefore contains the dynamical polarization of both partners automatically, together with neck formation, nucleon transfer and one-body dissipation \cite{SimenelUmar2018,Simenel2012}. Interaction potentials extracted from TDHF by the density-constrained method (DC-TDHF) \cite{Umar2006} and by the method of Washiyama and Lacroix \cite{Washiyama2008} show explicitly where the frozen-density picture ceases to apply. It was found in Ref. \cite{Washiyama2008} that at centre-of-mass energies well above the Coulomb barrier the microscopic potential coincides with the frozen-density one, whereas on approaching the barrier the potential becomes energy dependent, the barrier is lowered and its position moves outwards. This energy dependence is the signature of the rearrangement of the internal degrees of freedom which the frozen-density approximation omits.

These approaches establish that dynamical polarization is present and that it matters near the barrier, but they do not by themselves answer the question which is relevant for the DNS approach. In TDHF and DC-TDHF the polarization is inseparable from the other processes which occur simultaneously, and the quantity which is obtained is the potential of the combined system rather than the density change of one of the partners. The response of a given nucleus to the field of the other one, decomposed into its Coulomb and nuclear parts and expressed through the multipole moments which enter the DNS driving potential, is not directly accessible in that way. Microscopic studies of induced polarization have also been carried out in the linear-response and random-phase approximations, mainly for the electromagnetic response and for the polarization potential in elastic scattering \cite{Baltz1979,Alder1975}, but not with the nuclear part of the external field constructed from the same effective interaction that generates the DNS potential.

The present work addresses this gap. The density change of one nucleus produced by the static external field of its partner is calculated explicitly, the Coulomb and nuclear contributions are separated, and the nuclear part is obtained by double folding of the Migdal interaction in exactly the form in which it is used in the DNS approach, so that the correction is compatible with the potential which governs the relative motion. The result is expressed through the induced multipole moments and the polarization energy, which are the quantities entering the DNS driving potential and the equations of relative motion, and a quantitative criterion is established for the range of separations in which the description of one nucleus in the fixed field of the other remains meaningful.

The advantages and limitations of this approach with respect to the fully dynamical methods should be stated clearly. It is a static calculation with a frozen partner, so that neck formation, nucleon transfer and dissipation are absent, and the response is obtained at the level of the unperturbed particle-hole spectrum without the residual interaction. TDHF and DC-TDHF contain all of these effects and are therefore superior as descriptions of the collision. On the other hand, the present treatment isolates one well-defined quantity, the density change of a single nucleus in a specified external field, which the dynamical methods do not provide separately; it allows the Coulomb and nuclear parts of the polarization to be examined independently; and it uses the same effective interaction and the same geometry as the DNS calculations, so that the result can be carried over to them without a change of framework. It is intended as a source of microscopic corrections to the DNS entrance channel, not as a substitute for a dynamical description of the collision.

The purpose of the present work is to calculate the density change
\begin{equation}
\delta\rho(\bm r;R)=\rho(\bm r;R)-\rho_0(\bm r),
\label{eq:drho}
\end{equation}
where $\rho_0$ is the ground-state density of the isolated nucleus and $\rho(\bm r;R)$ is its density in the presence of the partner nucleus placed at the centre-to-centre distance $R$, and to obtain from it the induced multipole moments and the polarization energy. The single-particle problem is solved with the axially deformed Woods--Saxon potential of \'Cwiok \textit{et al.} \cite{Cwiok1987}, in which the external field of the partner has been included as an additional local term of the Hamiltonian. The nuclear part of the external field is calculated by double folding of the Migdal density-dependent effective interaction, which is the same interaction that is used to construct the nucleus-nucleus potential in the DNS approach \cite{AdamianDNS1996,Migdal1983}. The \Ca$+$\Pb~system has been chosen because both partners are doubly magic and spherical in their ground states, so that any deformation of the density is entirely induced by the partner, and because this system has been studied extensively as a reference case for quasifission and deep-inelastic processes. The restriction to spherical partners is deliberate. It removes the static deformation of the ground state from the problem and allows the induced polarization to be identified without ambiguity, which is necessary in order to establish the method and its limits. The formalism presented below is not restricted to spherical shapes, and the extension to statically deformed nuclei, where the orientation of the symmetry axes relative to the internuclear axis becomes an additional degree of freedom, is left to a subsequent work.

The choice of the effective interaction which generates the external nuclear field deserves particular attention. In the DNS approach the folded potential is used as the interaction energy between two nuclei, and its range is governed by the overlap of the two density distributions. If the same expression is replaced by a Woods--Saxon well whose radius is taken as the sum of the two nuclear radii, which is a natural parametrization of the nucleus-nucleus potential as a function of $R$, the resulting field acting on a single nucleon extends over a region much larger than the partner nucleus itself. Such a field does not represent the interaction of one nucleon with the nucleus $B$ and, as shown in Sec. III, it destroys the one-centre description at separations where the two nuclei are still well separated. The distinction between a nucleus-nucleus interaction radius and the range of a single-nucleon field is therefore essential in this type of calculation.

Two different treatments of the external field are compared. In the first one the field is added to the mean-field Hamiltonian and the eigenvalue problem is solved again, so that the response is obtained to all orders in the external field. In the second one the density change is evaluated in first-order perturbation theory, which is the standard linear-response expression. The comparison of the two provides a direct measure of the importance of the higher-order terms and, at the same time, an internal test of the calculations.

It should be stressed that the description of one nucleus in the fixed field of another one is meaningful only as long as the two nuclei retain their individuality. When the external field becomes strong enough to bind a nucleon in the region of the partner, the single-particle states of the combined system are no longer localized in one of the nuclei and the notion of the density of a single nucleus loses its meaning. The conditions under which this happens are examined quantitatively in the present work, and a simple criterion for the applicability of the one-centre description is proposed.

This paper is structured as follows: Sec. II presents the theoretical framework, the external field and the two methods used to obtain the density change; Sec. III discusses the range of validity of the one-centre picture; Sec. IV presents the results for the \Ca$+$\Pb~system. Finally, the conclusions are reported in Sec. V.

\section{Theoretical framework}

\subsection{Single-particle Hamiltonian of the isolated nucleus}

The single-particle states of the isolated nucleus are obtained from the axially symmetric Woods--Saxon Hamiltonian
\begin{equation}
\hat h_0=\hat t+V_{\rm WS}(\bm r)+V_{\rm so}(\bm r)+V_{\rm C}(\bm r),
\label{eq:h0}
\end{equation}
where $\hat t$ is the kinetic energy operator, $V_{\rm WS}$ is the central Woods--Saxon term, $V_{\rm so}$ is the spin-orbit term and $V_{\rm C}$ is the Coulomb potential of the nucleus itself, acting on protons only. The central term is written as
\begin{equation}
V_{\rm WS}(\bm r)=\frac{V_0}{1+\exp\left[\,\mathrm{dist}(\bm r,\Sigma)/a\,\right]},
\label{eq:ws}
\end{equation}
\begin{equation}
V_0=-V_{00}\left(1\pm\kappa\frac{N-Z}{A}\right),
\label{eq:ws2}
\end{equation}
where $\mathrm{dist}(\bm r,\Sigma)$ is the distance from the point $\bm r$ to the nuclear surface $\Sigma$, $a$ is the diffuseness parameter, and the upper and lower signs refer to protons and neutrons respectively. The parameters used in the present calculations are those of the universal parametrization, $V_{00}=40.6$ MeV, $r_0=1.24$ fm and $a=0.70$ fm for \Pb.

The eigenvalue problem for $\hat h_0$ is solved by diagonalization in the basis of the axially deformed harmonic oscillator, in cylindrical coordinates $(\rho,z,\varphi)$,
\begin{equation}
\phi_{\alpha}(\rho,z,\varphi)=
\varphi_{n_z}\!\left(\frac{z}{b_z}\right)
\varphi_{n_\rho}^{\Lambda}\!\left(\frac{\rho}{b_\rho}\right)
\frac{e^{i\Lambda\varphi}}{\sqrt{2\pi}}\,\chi_\sigma ,
\label{eq:basis}
\end{equation}
where $\varphi_{n_z}$ and $\varphi_{n_\rho}^{\Lambda}$ are the normalized Hermite and Laguerre functions, $\chi_\sigma$ is the spin function and the oscillator lengths are
\begin{equation}
b_z=\frac{\hbar c}{\sqrt{m c^2\,\hbar\omega_z}},\qquad
b_\rho=\frac{\hbar c}{\sqrt{m c^2\,\hbar\omega_\rho}} .
\label{eq:blength}
\end{equation}
The oscillator frequency is taken as $\hbar\omega_0=\lambda\cdot 41\,A^{-1/3}$ MeV with the basis scaling factor $\lambda=1.2$. The nuclear surface $\Sigma$ entering Eq. (\ref{eq:ws}) is parametrized by the multipole deformation parameters $\beta_2,\beta_3,\dots,\beta_6$, so that the mean field admits axially deformed shapes; in the present work all of them are set to zero, since both nuclei considered are spherical in their ground states. Since the Hamiltonian is invariant under rotations about the symmetry axis and under time reversal, the matrix is block diagonal in the projection $\Omega$ of the total angular momentum on the symmetry axis and in the parity $\pi$, and only the blocks with $\Omega>0$ have to be diagonalized. Every level obtained in this way is twofold degenerate, the two members of the Kramers doublet corresponding to $\pm\Omega$. The basis has been truncated at $N_{\rm max}=15$ oscillator shells, which gives 816 single-particle states for each isospin.

\subsection{External field of the partner nucleus}

The partner nucleus $B$ is placed on the symmetry axis of the nucleus $A$ under study, at $z=R$. This choice preserves the axial symmetry of the problem and keeps $\Omega$ a good quantum number. The distance from a point $(\rho,z)$ to the centre of $B$ is
\begin{equation}
d(\rho,z;R)=\sqrt{\rho^2+(z-R)^2}.
\label{eq:dist}
\end{equation}
The external field is the sum of a Coulomb and a nuclear part,
\begin{equation}
U_{\rm ext}(\bm r;R)=U^{(B)}_{\rm C}(\bm r;R)+U^{(B)}_{\rm N}(\bm r;R),
\label{eq:uext}
\end{equation}
the Coulomb part acting on protons only.

\subsubsection{Coulomb part}

The Coulomb field created by the charge distribution of $B$ is calculated in the uniformly charged sphere approximation,
\begin{equation}
U^{(B)}_{\rm C}(d)=
\begin{cases}
\dfrac{Z_B e^2}{d}, & d>R_B^{\rm ch},\\[2ex]
\dfrac{Z_B e^2}{2R_B^{\rm ch}}\left(3-\dfrac{d^2}{(R_B^{\rm ch})^2}\right), & d\le R_B^{\rm ch},
\end{cases}
\label{eq:ucoul}
\end{equation}
with $R_B^{\rm ch}=1.2\,A_B^{1/3}$ fm. At the separations considered here the whole volume of the nucleus $A$ lies outside the charge distribution of $B$, so that only the first line of Eq. (\ref{eq:ucoul}) is effective. The multipole expansion of the point charge field about the centre of $A$ gives
\begin{equation}
U^{(B)}_{\rm C}(\bm r)=\frac{Z_Be^2}{R}
\sum_{\lambda}\left(\frac{r}{R}\right)^{\lambda}P_\lambda(\cos\theta),
\label{eq:multipole}
\end{equation}
so that the term which drives the quadrupole polarization has the strength $Z_Be^2/(2R^3)$ and falls off as $R^{-3}$. This tidal behaviour is used below as a test of the calculations.

\subsubsection{Nuclear part}

The nuclear part of the external field is obtained by folding the density of the partner nucleus with the effective nucleon-nucleon interaction. The Migdal density-dependent zero-range force is used,
\begin{equation}
v_{\rm eff}(\bm r_1,\bm r_2)=C_0\left[f_{\rm ex}'+f_{\rm in}''\,
\frac{\rho(\bm r_1)}{\rho_0}\right]\delta(\bm r_1-\bm r_2),
\label{eq:migdal}
\end{equation}
where $\rho=\rho_A+\rho_B$ is the total nucleon density at the point of interaction, $\rho_0=0.17$ fm$^{-3}$ is the saturation density and $C_0=300$ MeV fm$^3$. The isospin dependence enters through
\begin{equation}
f_{\rm ex}'=f_{\rm ex}+f_{\rm ex}^{1}\,\xi,\qquad
f_{\rm in}''=f_{\rm in}+f_{\rm in}^{1}\,\xi-f_{\rm ex}',
\label{eq:migdalpar}
\end{equation}
with $\xi=\left[(2Z_A-A_A)/A_A\right]\left[(2Z_B-A_B)/A_B\right]$ and the parameter values $f_{\rm in}=0.09$, $f_{\rm ex}=-2.59$, $f_{\rm in}^{1}=0.42$ and $f_{\rm ex}^{1}=0.54$ \cite{Migdal1983,AdamianDNS1996}. These are the same interaction and the same parameter set that are used to construct the nucleus-nucleus potential in the DNS approach, so that the external field employed here is consistent with the potential which governs the relative motion. It should be noted that for the \Ca$+$\Pb~system the isospin factor $\xi$ vanishes because \Ca~has $N=Z$, and therefore $f_{\rm ex}'=-2.59$ and $f_{\rm in}''=2.68$.

Folding Eq. (\ref{eq:migdal}) with the two densities gives the nucleus-nucleus potential
\begin{equation}
V_{\rm N}(R)=C_0\!\int\! \rho_A(\bm r)\rho_B(\bm r-\bm R)
\left[f_{\rm ex}'+f_{\rm in}''\frac{\rho_A+\rho_B}{\rho_0}\right]d^3r .
\label{eq:vnn}
\end{equation}
The field acting on a single nucleon of $A$ is the functional derivative of Eq. (\ref{eq:vnn}) with respect to $\rho_A$,
\begin{equation}
U^{(B)}_{\rm N}(\bm r;R)=\frac{\delta V_{\rm N}}{\delta\rho_A(\bm r)}
=C_0\rho_0\,\hat\rho_B\!\left[f_{\rm ex}'+f_{\rm in}''
\left(2\hat\rho_A+\hat\rho_B\right)\right],
\label{aeq:unucl}
\end{equation}
where $\hat\rho_i=\rho_i/\rho_0$ and the term proportional to $\hat\rho_A$ is the rearrangement contribution arising from the density dependence of the force. The densities are taken in the two-parameter Fermi form with the effective radii used in the DNS calculations,
\begin{equation}
R_i=c_i\sqrt{\frac{R_p^2 Z_i+R_n^2 (A_i-Z_i)}{A_i}},
\label{eq:reff}
\end{equation}
with $R_p=1.237\left(1-0.157\delta_i-0.646/A_i\right)A_i^{1/3}$, $R_n=1.176\left(1+0.25\delta_i+2.806/A_i\right)A_i^{1/3}$, $\delta_i=(A_i-2Z_i)/A_i$, and diffuseness $a_0=0.54$ fm. The values obtained are $R=3.97$ fm for \Ca~and $R=7.03$ fm for \Pb. Since both nuclei considered here are spherical, the radii of Eq. (\ref{eq:reff}) do not depend on the direction. In the general case the surface of the partner is written as $R_i(\theta)=R_i\left[1+\sum_\lambda\beta_{\lambda i}Y_{\lambda 0}(\theta)\right]$ and the folding acquires a dependence on the orientation angles of the two symmetry axes with respect to the internuclear axis, in the same way as in the calculation of the nucleus-nucleus potential in the DNS approach \cite{Adamian2000}. The external field then loses its axial symmetry unless both nuclei are oriented along the internuclear axis, and $\Omega$ ceases to be a good quantum number for arbitrary orientations.

\begin{figure}[t]
\includegraphics[width=\columnwidth]{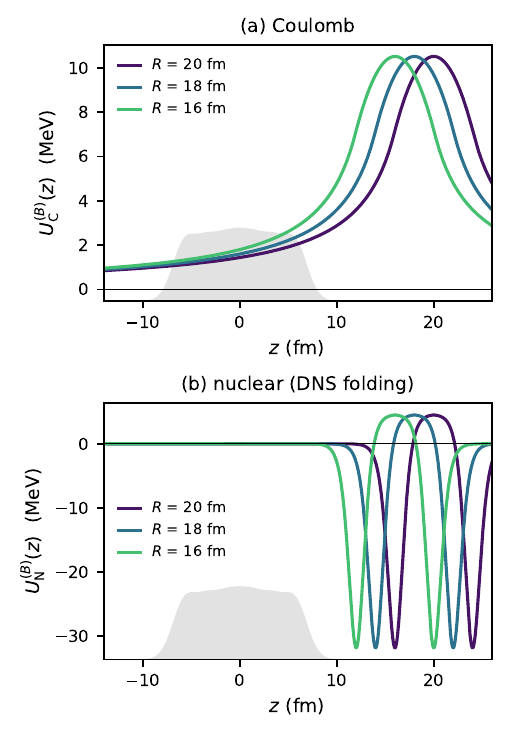}
\caption{(Color online) Coulomb (a) and nuclear (b) components of the external field created by \Ca~along the symmetry axis, for three centre-to-centre distances. The shaded area shows the ground-state density of \Pb~on an arbitrary scale. The nuclear component follows the density of the partner and does not extend into the volume of \Pb.}
\label{fig:fields}
\end{figure}

Because the force of Eq. (\ref{eq:migdal}) has zero range, the external field of Eq. (\ref{aeq:unucl}) follows the density of the partner directly and has no tail beyond it. This property is essential for the applicability of the whole approach and is discussed in Sec. III. It is also important to note that the Migdal force is repulsive at saturation density, $f_{\rm in}>0$, so that $U^{(B)}_{\rm N}$ is positive at the centre of the partner and reaches its largest attractive value, about $-32$ MeV, in the surface region of $B$. The behaviour of the two components of the external field along the symmetry axis is shown in Fig. \ref{fig:fields} for three separations, together with the density profile of \Pb.

\subsection{Non-perturbative treatment}

In the first method the external field is added to the mean-field Hamiltonian,
\begin{equation}
\hat h=\hat h_0+\lambda\,U_{\rm ext}(\bm r;R),
\label{eq:hfull}
\end{equation}
where the switching parameter $\lambda\in[0,1]$ is introduced in order to test the linearity of the response, and the eigenvalue problem is solved again in the same oscillator basis. The field of Eq. (\ref{eq:uext}) is evaluated on the Gauss--Hermite and Gauss--Laguerre mesh used for the matrix elements of the Woods--Saxon potential, so that no additional approximation is introduced in the construction of the Hamiltonian matrix.

Since the partner nucleus is located at $z=R$, the external field is not invariant under the reflection $z\to-z$, and parity is no longer a good quantum number. The blocks of the Hamiltonian matrix corresponding to the two parities of a given $\Omega$ are therefore merged and diagonalized together, so that only $\Omega$ remains a good quantum number in the perturbed problem. If this is not done, the calculation responds to the mirror image $U_{\rm ext}(-|z|)$ of the true field, and all odd multipole moments of $\delta\rho$ vanish identically.

\subsection{Linear response}

In the second method the density change is obtained in first order in $U_{\rm ext}$ from the unperturbed eigenstates,
\begin{equation}
\delta\rho(\bm r;R)=\sum_{ph}\frac{\langle p|U_{\rm ext}|h\rangle}
{\varepsilon_h-\varepsilon_p}\,\phi_p^{*}(\bm r)\phi_h(\bm r)+{\rm c.c.},
\label{eq:lr}
\end{equation}
where $h$ and $p$ label occupied and empty single-particle states of $\hat h_0$. The corresponding change of any one-body observable $\hat F$ is
\begin{equation}
\Delta F=2g\sum_{ph}\frac{\langle p|U_{\rm ext}|h\rangle\langle p|\hat F|h\rangle}
{\varepsilon_h-\varepsilon_p},
\label{eq:lrf}
\end{equation}
with $g=2$ the Kramers degeneracy. Both $U_{\rm ext}$ and the multipole operators are axially symmetric and spin independent, so that they conserve $\Omega$ and the sum in Eq. (\ref{eq:lrf}) runs inside each $\Omega$ block. The matrix elements are evaluated by Gauss--Legendre quadrature in the same oscillator basis. The residual interaction is not included, so that Eq. (\ref{eq:lr}) corresponds to the independent-particle response $\chi_0$; the random-phase approximation would replace it by $\delta\rho=(1-\chi_0\hat f)^{-1}\chi_0 U_{\rm ext}$ \cite{RingSchuck}.

The linear-response expression is built on the unperturbed states and therefore cannot become unstable, whatever the strength of the external field. Its accuracy, on the other hand, is limited to the region where the response is actually linear. The comparison of Eqs. (\ref{eq:lrf}) and (\ref{eq:hfull}) presented in Sec. IV shows where this condition is satisfied.

\subsection{Occupation numbers and observables}

The occupation numbers are obtained from the BCS equations with a constant pairing strength $G=18$ MeV fm$^3$. Since only the $\Omega>0$ blocks are computed, the number equation reads $N=2\sum_i v_i^2$, while the gap equation is a sum over Kramers doublets and carries no such factor. For the doubly magic nuclei considered here the pairing gap collapses and the occupation reduces to the sharp filling of the lowest $N/2$ levels. The density is then
\begin{equation}
\rho(\rho,z)=2\sum_i v_i^2\,|\psi_i(\rho,z)|^2 ,
\label{eq:dens}
\end{equation}
where the two $(\Lambda,\sigma)$ channels contained in each $\Omega$ block are summed incoherently, since they differ in the $e^{i\Lambda\varphi}$ factor and in the spinor component.

From $\delta\rho$ the following quantities are obtained,
\begin{eqnarray}
\delta A&=&\int\delta\rho\,d^3r\;\simeq\;0,\label{eq:obs1}\\
\Delta\langle r^2\rangle&=&\frac{1}{A}\int r^2\,\delta\rho\,d^3r,\label{eq:obs2}\\
\Delta Q_{20}&=&\int (2z^2-x^2-y^2)\,\delta\rho\,d^3r,\label{eq:obs3}\\
\Delta D_{1}&=&\int z\,\delta\rho\,d^3r,\label{eq:obs4}\\
\Delta E&=&\frac{1}{2}\int U_{\rm ext}\,\delta\rho\,d^3r,\label{eq:obs5}
\end{eqnarray}
the first of which serves as a check of the numerical accuracy. A positive $\Delta Q_{20}$ corresponds to a prolate elongation of the density along the internuclear axis and a negative one to an oblate compression.

\subsection{Numerical accuracy}

Several tests have been performed in order to control the numerical accuracy of the calculations. The densities are evaluated on a mesh covering $|z|\le16$ fm and $\rho\le13$ fm with $100\times70$ points, which is sufficient to conserve the integrated particle number to better than $10^{-4}$ for $A=208$; a smaller box truncates the tail of the density and loses about $0.7\%$ of the nucleons. The matrix elements entering Eq. (\ref{eq:lrf}) are evaluated with $220\times130$ Gauss--Legendre points, for which the orthonormality of the 816 single-particle states is satisfied to $7\times10^{-6}$. The quadrupole moment of the isolated \Pb, which must vanish for a closed-shell nucleus, is obtained as $0.04$ fm$^2$, that is, three orders of magnitude below the induced moments
discussed in Sec. IV.

The dependence of the results on the size of the oscillator basis has been examined by repeating the calculations with $N_{\rm max}=10$, 12 and 15. In the range of separations where the one-centre description is applicable the induced quadrupole moment is stable to within $2\%$. The behaviour outside this range is discussed in the next section.

\section{Validity of the one-centre description}

Before presenting the results it is necessary to establish the range of separations in
which the calculation of one nucleus in the field of the other is meaningful. The
oscillator basis of Eq. (\ref{eq:basis}) is centred on the nucleus $A$ and extends up to
$z_{\rm max}\simeq b_z\sqrt{2N_{\rm max}+3}=12.8$ fm for $N_{\rm max}=15$. If the
external field is attractive enough within this region, the variational solution places
amplitude in the well of the partner. The corresponding states are deeply bound, and
since the occupation is determined by the energy ordering, they become occupied. The
calculation then describes the combined system rather than the polarized nucleus $A$.

\begin{figure}[t]
\includegraphics[width=\columnwidth]{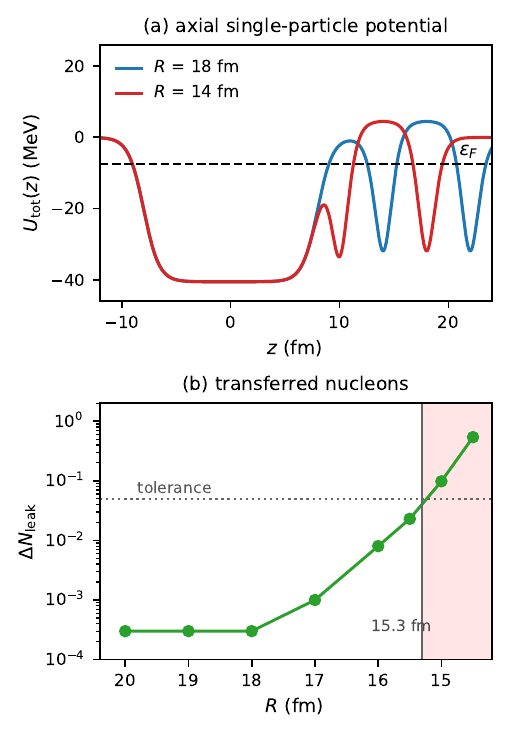}
\caption{(Color online) (a) Total single-particle potential of \Pb~along the symmetry axis in the
external field of \Ca. At $R=18$ fm a barrier rising above the Fermi energy separates the
two wells; at $R=14$ fm the wells have merged and no such barrier remains. The double minimum of the partner well reflects the repulsive character
of the Migdal force at saturation density. (b) Number of nucleons transferred beyond
$z_s=10$ fm as a function of the separation. The dotted line is the tolerance adopted in
this work and the shaded area marks the region in which the one-centre description fails.}
\label{fig:validity}
\end{figure}

Figure \ref{fig:validity}(a) shows the total single-particle potential along the symmetry
axis, that is, the Woods--Saxon well of \Pb~plus the external field of \Ca. For
$R\gtrsim16$ fm a barrier separates the two wells and rises above the Fermi level. Below
$R\simeq15$ fm this barrier disappears, the two wells merge, and the single-particle
states are no longer localized in one of the nuclei. The top of the barrier falls below
$\varepsilon_F$ at $R\simeq15.3$ fm.

The onset of this behaviour is monitored by the number of nucleons which appear beyond
the surface of the nucleus $A$,
\begin{equation}
\Delta N_{\rm leak}=\int_{z>z_s}\rho(\bm r;R)\,d^3r-\int_{z>z_s}\rho_0(\bm r)\,d^3r ,
\label{eq:leak}
\end{equation}
with $z_s=10$ fm for \Pb. This quantity is shown in Fig. \ref{fig:validity}(b). It stays
below $10^{-3}$ down to $R=17$ fm, reaches $2\times10^{-2}$ at $R=15.5$ fm and then rises by
more than an order of magnitude per fm, crossing the tolerance adopted here at
$R\simeq15.3$ fm. The two criteria are independent, the first being a property of the
potential alone and the second of the resulting wave functions, and they locate the
breakdown at the same separation.

A calculation which has entered this regime does not converge with the size of the basis.
This has been verified for a case in which the external field is deliberately taken in the
form of a wide Woods--Saxon well of depth 50 MeV and radius $1.2(A_A^{1/3}+A_B^{1/3})$,
that is, with the geometry of a nucleus-nucleus interaction potential rather than of a
single-nucleon field. At $R=20$ fm the induced quadrupole moment grows monotonically from
248 to 687 and 2466 fm$^2$ as $N_{\rm max}$ is increased from 10 to 12 and 15, whereas at
$R=26$ fm it remains stable at $-3.1$ fm$^2$ for the same three basis sizes. The absence of
convergence is the signature of a variational collapse: a larger basis reaches further into
the well of the partner, lowers the energy and transfers more nucleons. A result obtained
in this regime is therefore not defined independently of the basis, whereas in the range
retained in this work the induced moments are stable to within $2\%$ over the same range of
$N_{\rm max}$.

\begin{figure}[t]
\includegraphics[width=\columnwidth]{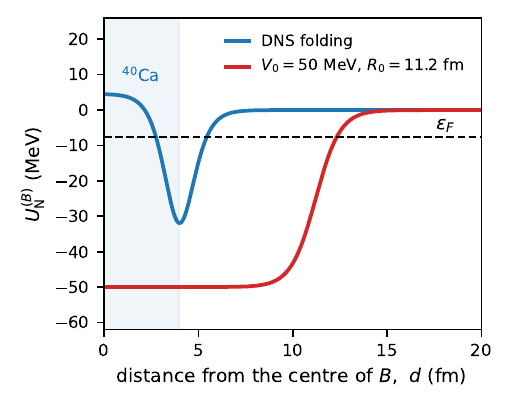}
\caption{(Color online) Nuclear part of the external field of \Ca~as a function of the distance to its
centre, obtained from the folding of Eq. (\ref{aeq:unucl}) and from a Woods--Saxon well
whose radius is taken as the sum of the two nuclear radii,
$R_0=1.2(A_A^{1/3}+A_B^{1/3})=11.2$ fm. The shaded band indicates the extent of the
\Ca~density and the dashed line the Fermi energy of \Pb.}
\label{fig:fieldcomp}
\end{figure}

The geometry assumed for the external nuclear field determines where this limit lies.
Figure \ref{fig:fieldcomp} compares the folded field of Eq. (\ref{aeq:unucl}) with a
Woods--Saxon well of depth 50 MeV and radius $1.2(A_A^{1/3}+A_B^{1/3})=11.2$ fm. The
folded field is confined to the density of the partner and falls below the Fermi energy
already at $d\simeq6$ fm, whereas the wide well remains 50 MeV deep out to 10 fm and
therefore reaches the surface of \Pb~even when the two centres are 20 fm apart. Applying
the criterion of Eq. (\ref{eq:criterion}) to the two fields gives $R>15.3$ fm for the
folded one and $R>22.3$ fm for the wide well, so that the two limits differ by 7 fm. It is
therefore the range of the external field, and not its depth, which controls the
applicability of the one-centre picture.

These observations lead to a simple criterion. The one-centre description is applicable
as long as the external field is unable to bind a nucleon anywhere the nucleus $A$ has
appreciable density,
\begin{equation}
\left|U_{\rm ext}(z_s)\right|<\left|\varepsilon_F\right| ,
\qquad z_s\simeq R_A+2a ,
\label{eq:criterion}
\end{equation}
where $\varepsilon_F$ is the Fermi energy of the nucleus $A$. For the \Ca$+$\Pb~system
with the folded field of Eq. (\ref{aeq:unucl}) this gives $R>15.3$ fm, in agreement with
the disappearance of the barrier in Fig. \ref{fig:validity}(a) and with the rise of
$\Delta N_{\rm leak}$ in Fig. \ref{fig:validity}(b). All results presented below have been
obtained in this range.

It is important to note that the limit is a physical one and not a numerical artifact.
The touching configuration of \Ca$+$\Pb~corresponds to $R\simeq12.1$ fm, so that the
interval between 12.1 and 15.3 fm is not accessible to any one-centre frozen calculation.
In this region the two nuclei genuinely share single-particle states, and a two-centre
basis is required.

\section{Results and discussion}

\subsection{Ground-state properties of the isolated nuclei}

\begin{figure}[t]
\includegraphics[width=\columnwidth]{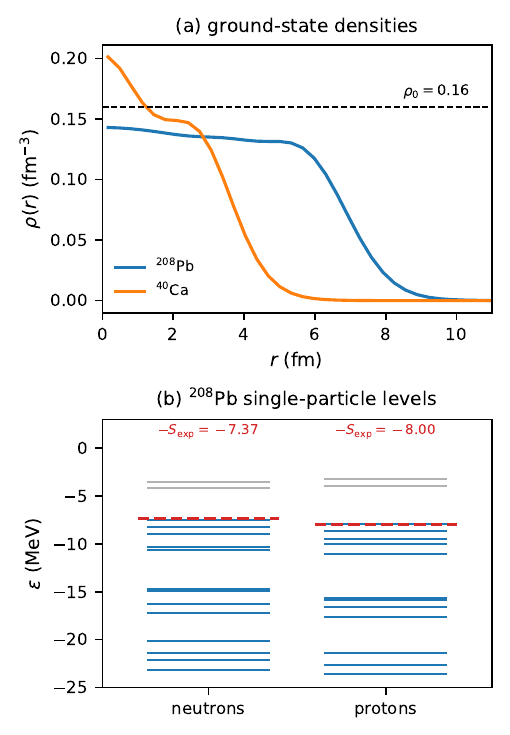}
\caption{(Color online) (a) Radial density distributions of the isolated \Pb~and \Ca~nuclei obtained in
the present calculation. The dashed line indicates the saturation density.
(b) Single-particle levels of \Pb~near the Fermi surface; occupied levels are shown by
full lines and the lowest empty ones by grey lines. The dashed lines give the experimental
neutron and proton separation energies with the opposite sign.}
\label{fig:baseline}
\end{figure}

The quality of the unperturbed mean field determines the reliability of the calculated
polarization, and therefore the ground-state properties of the two nuclei have been
examined first. The results are collected in Table \ref{tab:gs} and the density profiles
and single-particle levels are shown in Fig. \ref{fig:baseline}.

\begin{table}[h]
\caption{Ground-state properties of the isolated nuclei obtained in the present
calculation, in comparison with the experimental data. The root-mean-square radii are the
measured charge radii of Ref. \cite{Angeli2013}, the central densities are the values
deduced from elastic electron scattering \cite{deVries1987} converted to nucleon density,
and the Fermi energies are compared with the one-nucleon separation energies of
Ref. \cite{Wang2021} taken with the opposite sign.}
\label{tab:gs}
\begin{ruledtabular}
\begin{tabular}{lcccc}
 & \multicolumn{2}{c}{\Pb} & \multicolumn{2}{c}{\Ca} \\
 & calc. & exp. & calc. & exp. \\ \hline
$A$ (integrated)         & 207.94 & 208    & 39.97 & 40 \\
$\langle r^2\rangle^{1/2}$ (fm) & 5.83 & 5.50 & 3.43 & 3.48 \\
$\rho(0)$ (fm$^{-3}$)    & 0.143  & $\sim$0.16 & 0.201 & $\sim$0.17 \\
$\varepsilon_F^{(n)}$ (MeV) & $-7.54$ & $-7.37$ & $-15.64$ & $-15.64$ \\
$\varepsilon_F^{(p)}$ (MeV) & $-7.94$ & $-8.00$ & $-6.60$  & $-8.33$ \\
\end{tabular}
\end{ruledtabular}
\end{table}

The integrated particle numbers reproduce the input values to better than $0.1\%$. The
root-mean-square radius of \Ca~agrees with the measured charge radius of
$3.478$ fm \cite{Angeli2013} within $1.5\%$, while that of \Pb~comes out about $6\%$ larger
than the measured value of $5.501$ fm \cite{Angeli2013} with the parameter set used here.
The central density of \Pb~is somewhat below the saturation value of nuclear matter and
that of \Ca~is above it. Both features are consistent with the charge distributions
obtained from elastic electron scattering \cite{deVries1987}, which show a central
depression in \Pb~and a central bump in \Ca, the latter being characteristic of a light
$N=Z$ nucleus and enhanced in a mean-field description which does not include short-range
correlations. The last occupied single-particle level is to be compared with the
one-nucleon separation energy taken with the opposite sign, $\varepsilon_F=-S$. For \Pb~
the last occupied neutron level lies at $-7.54$ MeV against $-S_n=-7.368$ MeV
\cite{Wang2021}, and the corresponding proton level at $-7.94$ MeV against
$-S_p=-8.004$ MeV. For \Ca~the neutron Fermi energy of $-15.64$ MeV reproduces
$-S_n=-15.641$ MeV exactly, while the proton level at $-6.60$ MeV lies about $1.7$ MeV
above $-S_p=-8.328$ MeV, that is, it is less bound than the measured value; this deviation
of the universal parametrization for the lighter partner has been kept in mind when
interpreting the results for \Ca.

\begin{figure}[b]
\includegraphics[width=0.85\columnwidth]{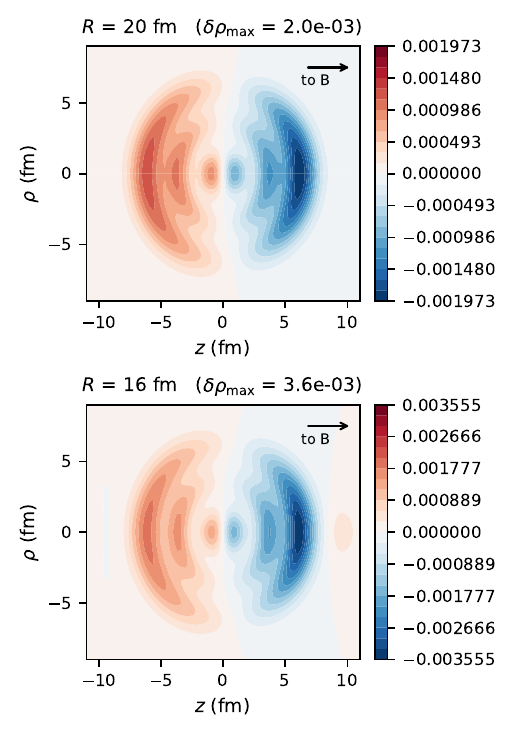}
\caption{(Color online) Density change $\delta\rho(z,\rho)$ induced in \Pb~by the presence of \Ca~at
$R=20$ fm (upper panel) and $R=16$ fm (lower panel). The partner nucleus is located on the positive
$z$ axis, as indicated by the arrow. Red and blue correspond to an increase and to a
decrease of the density respectively.}
\label{fig:drho_maps}
\end{figure}

\subsection{Structure of the polarization density}

The density change induced in \Pb~by \Ca~is shown in Fig. \ref{fig:drho_maps} for two
separations. The polarization is concentrated in the surface region of the nucleus. At
$R=16$ fm the fraction of $|\delta\rho|$ contained between $r=4$ and $r=9$ fm amounts to
about three quarters of the total, and the weighted mean radius of $|\delta\rho|$ is
$5.3$ fm, close to the root-mean-square radius of the nucleus itself. The interior of the
nucleus is almost unaffected, which is a consequence of the fact that the external field
is smooth over the nuclear volume and that the deeply bound orbitals are separated from
the empty ones by a large energy denominator.

\begin{figure}[h]
\includegraphics[width=\columnwidth]{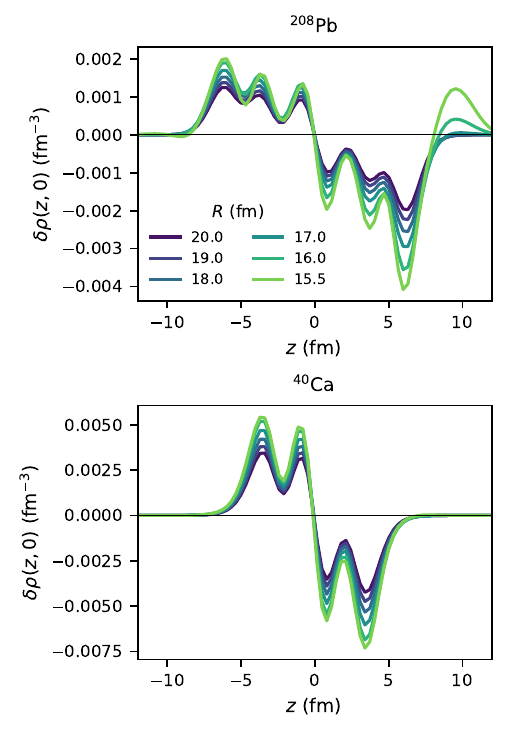}
\caption{(Color online) Density change along the symmetry axis for \Pb~(upper panel) and \Ca~(lower panel) at the
separations indicated. The partner nucleus is on the positive $z$ side.}
\label{fig:drho_axis}
\end{figure}

The pattern of Fig. \ref{fig:drho_maps} shows an accumulation of density on the side
opposite to the partner and a depletion on the side facing it. This is the expected
response of a positively charged system to the repulsive field of another positive charge.
The effect is essentially a displacement of the proton distribution: the calculated dipole
moment of the total density change at $R=20$ fm is
$\Delta D_1=-4.26$ fm, and the value obtained from the protons alone is $-4.26$ fm, so
that the neutrons do not contribute. The corresponding shift of the centre of the proton
distribution with respect to the centre of the mean field is
$\Delta D_1/Z=-0.052$ fm. It should be noted that the mean field of the nucleus $A$ is
held fixed at the origin in the present treatment, so that this dipole moment describes
the displacement of the charge distribution relative to the mean-field centre and should
not be identified with an intrinsic $E1$ moment of a freely recoiling nucleus.

The behaviour of $\delta\rho$ along the symmetry axis is presented in
Fig. \ref{fig:drho_axis} for both partners and for all separations considered. The
amplitude of the polarization grows monotonically as the two nuclei approach, and the
shape of the curves changes little, which indicates that the response remains dominated
by the same few particle-hole configurations over the whole range.

\subsection{Quadrupole polarization}

\begin{figure}[t]
\includegraphics[width=\columnwidth]{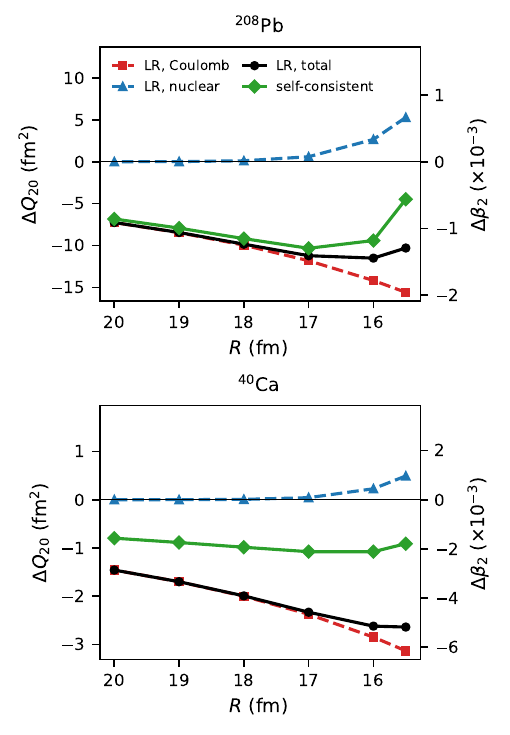}
\caption{(Color online) Induced quadrupole moment of \Pb~(upper panel) and \Ca~(lower panel) as a function of the
centre-to-centre distance. The full symbols connected by the solid green line show the
result of the re-diagonalization, and the black line the total linear-response result. The
dashed red and blue lines give the Coulomb and nuclear contributions to the
linear-response value separately. The right-hand scale gives the corresponding induced
quadrupole deformation parameter of Eq. (\ref{eq:beta}).}
\label{fig:dq20}
\end{figure}

The induced quadrupole moments are given in Table \ref{tab:results} and shown in
Fig. \ref{fig:dq20} together with the linear-response results and their decomposition into
the Coulomb and nuclear contributions.

\begin{table*}[t]
\caption{Induced quadrupole moment, mean-square radius change, dipole moment and
polarization energy of \Pb~and \Ca~as functions of the centre-to-centre distance. The
last column gives the number of nucleons which appear beyond $z_s=10$ fm and serves as a
measure of the reliability of the one-centre description.}
\label{tab:results}
\begin{ruledtabular}
\begin{tabular}{cddddc}
$R$ (fm) & \multicolumn{1}{c}{$\Delta Q_{20}$ (fm$^2$)}
         & \multicolumn{1}{c}{$\Delta\langle r^2\rangle$ (fm$^2$)}
         & \multicolumn{1}{c}{$\Delta D_1$ (fm)}
         & \multicolumn{1}{c}{$\Delta E$ (MeV)}
         & $\Delta N_{\rm leak}$ \\ \hline
\multicolumn{6}{c}{\Pb} \\ \hline
20.0 & -6.83  & 0.0009 & -4.26 & -0.160 & $<10^{-3}$ \\
19.0 & -7.92  & 0.0011 & -4.71 & -0.197 & $<10^{-3}$ \\
18.0 & -9.17  & 0.0016 & -5.24 & -0.246 & $<10^{-3}$ \\
17.0 & -10.33 & 0.0030 & -5.79 & -0.313 & 0.001 \\
16.0 & -9.40  & 0.0098 & -6.34 & -0.430 & 0.008 \\
15.5 & -4.47  & 0.0230 & -6.30 & -0.610 & 0.023 \\ \hline
\multicolumn{6}{c}{\Ca} \\ \hline
20.0 & -0.797 & 0.0102 & -2.21 & -0.329 & $<10^{-3}$ \\
19.0 & -0.887 & 0.0125 & -2.42 & -0.404 & $<10^{-3}$ \\
18.0 & -0.985 & 0.0155 & -2.72 & -0.500 & $<10^{-3}$ \\
17.0 & -1.078 & 0.0196 & -3.05 & -0.626 & $<10^{-3}$ \\
16.0 & -1.078 & 0.0264 & -3.39 & -0.792 & 0.001 \\
15.5 & -0.912 & 0.0325 & -3.57 & -0.894 & 0.002 \\
\end{tabular}
\end{ruledtabular}
\end{table*}

The induced quadrupole moment is negative for both partners over the whole range of
separations, that is, the density is compressed along the internuclear axis and the
resulting shape is oblate. This may appear surprising at first sight, since the nuclear
attraction of the partner acts in the direction of an elongation, but the decomposition
shown in Fig. \ref{fig:dq20} makes the origin of the effect clear. The quadrupole
component of the Coulomb field, Eq. (\ref{eq:multipole}), raises the potential at the
poles of the nucleus relative to its equator, so that the protons are pushed towards the
equatorial plane. A repulsive tidal field compresses, whereas an attractive one, such as
the gravitational tide, elongates. The nuclear contribution is indeed positive but it is
much smaller in the region where the calculation is valid.

The relative importance of the two contributions changes rapidly with the separation. At
$R=20$ fm the nuclear part of $\Delta Q_{20}$ amounts to $0.004$ fm$^2$ against
$-7.25$ fm$^2$ from the Coulomb part, that is, to less than $0.1\%$. At $R=17$ fm it has
grown to $0.60$ fm$^2$, and at $R=15.5$ fm it reaches $5.29$ fm$^2$, about a third of the
Coulomb value. This rapid growth is a direct consequence of the zero range of the Migdal
force: the external nuclear field is significant only where the density of the partner
is significant, and the overlap of the tail of \Pb~with the surface of \Ca~grows very fast
once the two surfaces approach each other.

It should be stressed that the oblate character found here is a property of the range of
separations in which the Coulomb tidal field dominates. The nuclear contribution grows
exponentially as the surfaces approach, with an e-folding length of $0.62$ fm obtained from
the present results, which is close to the diffuseness of the density distributions. An
extrapolation of this behaviour together with the $R^{-3}$ Coulomb term indicates a change
of sign near $R\simeq15$ fm, that is, just below the range accessible to the present
calculation. Since the Coulomb barrier of the \Ca$+$\Pb~system is located near
$R\simeq13$ fm, a prolate polarization is to be expected in the barrier region itself, in
agreement with the elongation reported in dynamical models of fusion. The two pictures
correspond to different ranges of the separation rather than to a disagreement.

The competition of the two contributions produces a turnover in the magnitude of the
polarization. For \Pb~the modulus of $\Delta Q_{20}$ increases from $6.83$ fm$^2$ at
$R=20$ fm to $10.33$ fm$^2$ at $R=17$ fm and then decreases again to $4.47$ fm$^2$ at
$R=15.5$ fm, the maximum being reached at $R\simeq17$ fm. For \Ca~the same behaviour is
observed with the maximum at $R\simeq16$--$17$ fm. It is important to note that no change
of sign takes place within the range of separations where the one-centre description is
valid; the cancellation between the Coulomb and nuclear tidal fields remains partial down
to $R=15.5$ fm.

\begin{figure}[b]
\includegraphics[width=0.90\columnwidth]{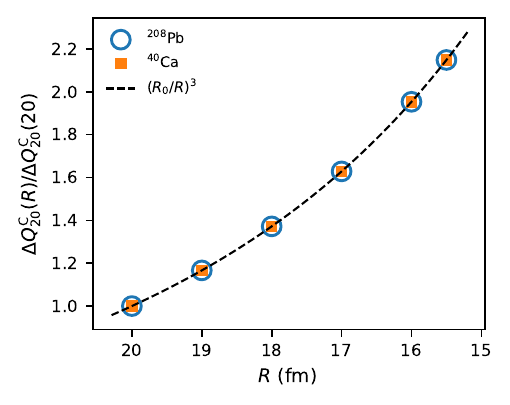}
\caption{(Color online) Coulomb contribution to the induced quadrupole moment, normalized to its value at
$R=20$ fm, in comparison with the $R^{-3}$ dependence expected for a tidal field. The
results for the two nuclei coincide, and \Pb~is therefore drawn with an open symbol.}
\label{fig:scaling}
\end{figure}

The Coulomb part of the response provides a test of the calculations. According to
Eq. (\ref{eq:multipole}) the quadrupole driving term scales as $R^{-3}$, and if the
response is linear the induced moment must follow the same law.
Figure \ref{fig:scaling} shows that the calculated Coulomb contribution reproduces the
$R^{-3}$ dependence for both nuclei over the whole range, with deviations below the
thickness of the line.

\subsection{Induced deformation parameter}

It is convenient to express the induced quadrupole moment through the dimensionless
deformation parameter, defined in the usual way as
\begin{equation}
\beta_2=\frac{4\pi}{3AR_0^2}\int\rho(\bm r)\,r^2Y_{20}(\theta)\,d^3r,
\qquad R_0=1.2A^{1/3},
\label{eq:beta}
\end{equation}
which is the convention in which the tabulated ground-state deformations are given
\cite{Raman2001,Moller1995}, so that the change induced by the partner is
$\Delta\beta_2=\sqrt{5\pi}\,\Delta Q_{20}/(3AR_0^2)$ with $\Delta Q_{20}$ of
Eq. (\ref{eq:obs3}). The values obtained in the two treatments are given in
Table \ref{tab:beta}.

\begin{table*}[t]
\caption{Induced quadrupole deformation parameter $\Delta\beta_2$ (in units of
$10^{-3}$) of the two partners, obtained by re-diagonalization (SC) and in first-order
perturbation theory (LR), and the ratio of the two.}
\label{tab:beta}
\begin{ruledtabular}
\begin{tabular}{cdddddd}
 & \multicolumn{3}{c}{\Pb} & \multicolumn{3}{c}{\Ca} \\
\multicolumn{1}{c}{$R$ (fm)}
 & \multicolumn{1}{c}{SC} & \multicolumn{1}{c}{LR} & \multicolumn{1}{c}{ratio}
 & \multicolumn{1}{c}{SC} & \multicolumn{1}{c}{LR} & \multicolumn{1}{c}{ratio} \\ \hline
20.0 & -0.858 & -0.910 & 0.94 & -1.562 & -2.859 & 0.55 \\
19.0 & -0.994 & -1.060 & 0.94 & -1.739 & -3.333 & 0.52 \\
18.0 & -1.152 & -1.234 & 0.93 & -1.932 & -3.908 & 0.49 \\
17.0 & -1.297 & -1.408 & 0.92 & -2.113 & -4.572 & 0.46 \\
16.0 & -1.180 & -1.444 & 0.82 & -2.113 & -5.139 & 0.41 \\
15.5 & -0.561 & -1.293 & 0.43 & -1.787 & -5.176 & 0.35 \\
\end{tabular}
\end{ruledtabular}
\end{table*}

The induced deformations are small, of the order of $10^{-3}$, which justifies a
posteriori the treatment of the polarization as a perturbation of the ground state. It is
important to note that although the induced quadrupole moment of \Pb~is about eight times
larger than that of \Ca~in absolute value, the deformation parameter behaves in the
opposite way: $\Delta\beta_2$ of the light partner is roughly twice that of the heavy one
at every separation. The reason is the normalization by $AR_0^2$ in Eq. (\ref{eq:beta}),
which is about sixteen times smaller for \Ca. The lighter nucleus is therefore the more strongly
polarized of the two in relative terms, both because it is softer and because the field
of \Pb~is four times stronger than the field of \Ca. This is relevant for the DNS
description, in which the deformation parameters of the two fragments enter the driving
potential on an equal footing.

\subsection{Comparison with the linear-response treatment}

The two treatments of the external field are compared in Fig. \ref{fig:dq20}. For \Pb~the
ratio of the re-diagonalized to the first-order value is $0.94$ at $R=20$ fm and decreases
slowly to $0.92$ at $R=17$ fm. In this interval the two calculations agree within less than
$10\%$, which shows that the response of the heavy partner to the field of \Ca~is
essentially linear. Below $R=17$ fm the agreement deteriorates rapidly and the ratio falls
to $0.43$ at $R=15.5$ fm. This is not a failure of either method but a consequence of the
partial cancellation discussed above: the total is a difference of two larger quantities,
and a small absolute discrepancy between them produces a large relative one.

For \Ca~the ratio is close to $0.55$ already at $R=20$ fm. The deviation from unity is
genuine and reflects the saturation of the response of the lighter partner to the much
stronger field of \Pb. This has been verified by repeating the calculation with the
switching parameter $\lambda$ of Eq. (\ref{eq:hfull}). For \Pb~the ratio of the
re-diagonalized to the first-order value is $0.990$ at $\lambda=0.1$, $0.981$ at
$\lambda=0.3$ and $0.942$ at $\lambda=1$, while for \Ca~the corresponding values are
$0.952$, $0.860$ and $0.547$. Both nuclei therefore converge to the linear-response result
in the limit of a weak field, and the departure at full strength measures the higher-order
terms. Since the charge of \Pb~is four times larger than that of \Ca, the field acting on
the lighter partner is correspondingly stronger and the higher-order terms are more
important.

\begin{figure}[t]
\includegraphics[width=\columnwidth]{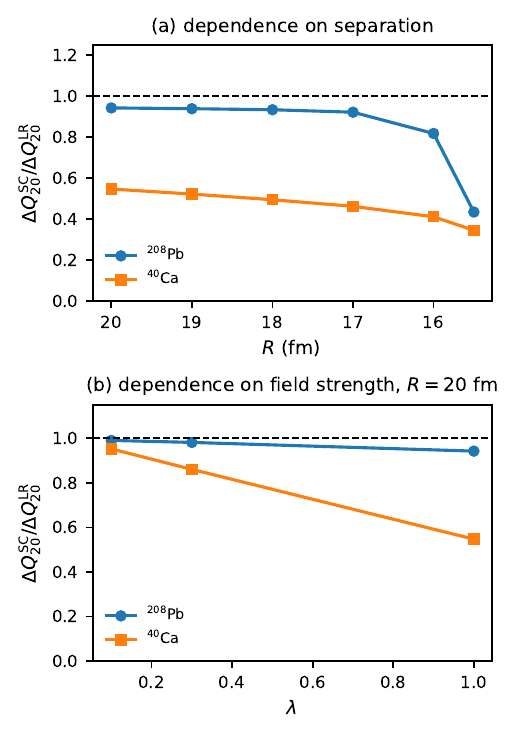}
\caption{(Color online) Ratio of the induced quadrupole moment obtained by re-diagonalization to its
first-order value, (a) as a function of the centre-to-centre distance at full strength of
the external field, and (b) as a function of the switching parameter $\lambda$ of
Eq. (\ref{eq:hfull}) at $R=20$ fm for the Coulomb field alone.}
\label{fig:method}
\end{figure}

The ratio of the two treatments is shown in Fig. \ref{fig:method} as a function of the
separation and of the switching parameter. Panel (b) demonstrates that the two methods
converge to the same result when the external field is switched off, which is the expected
behaviour and provides an internal consistency check of the two independent
implementations. The departure from unity at $\lambda=1$ measures the contribution of the
terms beyond first order, which is $6\%$ for \Pb~and $45\%$ for \Ca.

\subsection{Polarization energy and radius change}

\begin{figure}[t]
\includegraphics[width=\columnwidth]{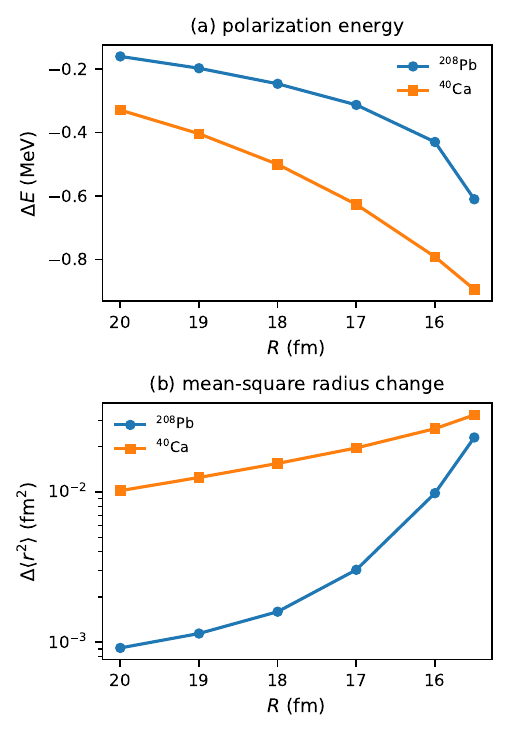}
\caption{(Color online) Polarization energy (a) and change of the mean-square radius (b) of the two
partners as functions of the centre-to-centre distance.}
\label{fig:energy}
\end{figure}

The polarization energy of Eq. (\ref{eq:obs5}) and the change of the mean-square radius
are shown in Fig. \ref{fig:energy}. The polarization energy is negative for both partners,
as it must be for an induced response, and grows in magnitude as the separation decreases.
At $R=15.5$ fm it amounts to $-0.61$ MeV for \Pb~and to $-0.89$ MeV for \Ca.

It is important to note that this energy is not a measure of the quadrupole polarization
which is the subject of the preceding subsections. Inserting the multipole expansion of
Eq. (\ref{eq:multipole}) into Eq. (\ref{eq:obs5}) gives
\begin{equation}
\Delta E\simeq\frac{Z_Be^2}{2R^2}\,\Delta D_1+\frac{Z_Be^2}{2R^3}\,\frac{\Delta Q_{20}}{2},
\label{eq:demult}
\end{equation}
and the two terms reproduce the calculated values to better than $2\%$ at $R\ge18$ fm. The
first term accounts for $95$ to $96\%$ of $\Delta E$ for \Pb~and for $99\%$ for \Ca. The
polarization energy is therefore a dipole quantity, and the quadrupole deformation
discussed above contributes only a few percent of it.

The dipole response obtained here is not screened, since the residual particle-hole
interaction is not included in Eq. (\ref{eq:hfull}). Its size can be tested against the
measured static dipole polarizability, which is the one quantity of the present
calculation for which direct experimental data exist. Evaluating
\begin{equation}
\alpha_D=2g\sum_{ph}\frac{|\langle p|\hat D_z|h\rangle|^2}{\varepsilon_p-\varepsilon_h}
\label{eq:alphaD}
\end{equation}
for the isolated nuclei, with the dipole operator taken with the full proton charge, gives
$85.7$ fm$^3$ for \Pb~and $10.9$ fm$^3$ for \Ca. These values agree with those extracted
from $\Delta D_1$ and the uniform part of the external field, $85$ and $11$ fm$^3$, which
are moreover almost independent of $R$; the two determinations are independent and their
agreement confirms the linearity of the response.

Part of this response is spurious. A uniform field acting on the protons of a nucleus whose
mean field is anchored at the origin displaces the whole system against an artificial
restoring force, and this centre-of-mass motion is not a polarization. It is removed in the
standard way by using the effective charges $e_p=Ne/A$ and $e_n=-Ze/A$ in
Eq. (\ref{eq:alphaD}), which gives $57.3$ fm$^3$ for \Pb~and $5.26$ fm$^3$ for \Ca. The
remaining values exceed the measured $20.1\pm0.6$ fm$^3$ \cite{Tamii2011} and
$1.92\pm0.17$ fm$^3$ \cite{Fearick2023} by factors of $2.9$ and $2.7$ respectively. It is
important to note that these two factors are equal to within the accuracy of the
comparison, although the two nuclei differ by a factor five in mass. The overestimate is
therefore a systematic property of the unperturbed particle-hole response and not a
numerical deficiency: the isovector restoring force which shifts the dipole strength up
into the giant resonance is absent, and it is known to reduce the inverse energy weighted
sum rule by a factor of this order. The values of $\Delta E$ quoted above should
accordingly be regarded as upper bounds, the screened estimates being smaller by
approximately the same factor.

The change of the mean-square radius is positive but very small, and it is almost entirely
a consequence of the displacement of the proton distribution rather than of a breathing of
the density. A rigid shift $d_p=|\Delta D_1|/Z$ of the protons raises $\langle r^2\rangle$
by $(Z/A)d_p^2$, which accounts for the whole of the calculated value for \Pb~and for about
$60\%$ of it for \Ca.
The effect is very small for \Pb, where $\Delta\langle r^2\rangle$ reaches only
$0.023$ fm$^2$ at $R=15.5$ fm, corresponding to an increase of the root-mean-square radius
by $0.002$ fm. For \Ca~the relative effect is larger, $0.033$ fm$^2$ at the same
separation, because of the smaller size of the nucleus and of the stronger field of the
heavy partner. The particle number is conserved to better than $10^{-3}$ in all cases,
which confirms the numerical accuracy of the calculations.

\subsection{Relevance for the equations of relative motion}

In the DNS approach the polarization of the fragment densities is not included in the
equations of relative motion. The nucleus-nucleus potential is constructed from the
ground-state densities of the two nuclei, which are kept frozen during the approach, and
the deformation parameters entering the driving potential are the static ones taken from
the tables. This assumption is a natural first approximation, but it has not been verified
microscopically, and the size of the correction which it neglects is not known beforehand.
One of the purposes of the present calculation is to provide that check.

The magnitudes obtained here can therefore be placed directly against the ingredients of
the equations of relative motion \cite{AdamianDNS1996,Adamian1997,Nasirov2005, Kayumov2025}. In that
framework the radial coordinate obeys $\mu(R)\ddot R=-\partial V/\partial R-\gamma_R(R)\dot R$, and
the density polarization can in principle enter through the interaction potential $V(R)$,
through the radial friction coefficient $\gamma_R$, through the coordinate-dependent
inertia $\mu(R)$, and through the single-particle spectrum which determines the nucleon
transfer rates. Of these, only the first is affected at a level which the present
calculation can quantify. The polarization energy of Eq. (\ref{eq:obs5}) is the
second-order correction to the frozen-density folding potential, it is negative by
construction, and it therefore acts as an additional attraction. What matters for the
trajectory is not $\Delta E$ itself but its radial derivative, which between $R=17$ and
$R=15.5$ fm amounts to about $0.4$ MeV/fm, that is, less than four percent of the Coulomb
force at the same separations. The radial friction coefficient is proportional to the
squared radial gradients of the single-particle coupling matrix elements
\cite{Adamian1997} and therefore depends on the density change only at second order; with
a relative change of the surface density of the order of $10^{-3}$ the resulting change of
$\gamma_R$ is negligible in comparison with the uncertainty associated with the one-body
treatment of the dissipation \cite{Randrup1984}. The mass parameter $\mu(R)$ is governed by
the overlap integral of the two density distributions and is unaffected at the separations
considered here, where that overlap vanishes.

The induced quadrupole deformations, $|\Delta\beta_2|\lesssim2\times10^{-3}$, are small in
comparison with the accuracy to which the static deformation parameters entering the
driving potential are known. The tabulated $\beta_2$ values used in such calculations
\cite{Raman2001,Moller1995} carry uncertainties of the order of $10^{-2}$, and the
discretization of the grids over the orientation angles of the symmetry axes corresponds to
still larger variations of the interaction potential. The quadrupole polarization computed
here is therefore not resolvable as a correction to the driving potential, and the frozen
deformation treatment of the fragments remains adequate for that purpose. It may be noted
that a polarization of the same sign is already generated in trajectory calculations in
which the quadrupole amplitudes are propagated as driven oscillators, since the Coulomb
driving term $\partial V_{\rm C}/\partial\beta_2$ is positive and drives the amplitude
towards oblate values.

The physical content of these numbers for the entrance channel may be stated as follows.
The polarization energy is an additional attraction which is absent from the folded
potential, so that the true adiabatic potential lies below the frozen-density one at every
separation. Since $|\Delta E|$ increases as the nuclei approach, the correction is not a
constant shift but a steepening of the potential, which lowers the barrier and moves its
position slightly outwards. This is the same qualitative behaviour as the energy dependence
of the microscopic potentials extracted from TDHF near the barrier
\cite{Umar2006,Washiyama2008}. The induced density change acts in the same direction
through a second channel: the redistribution of density towards the surface facing away
from the partner reduces the overlap of the two distributions at a given $R$, which weakens
the attractive folding term, but this shape effect is an order of magnitude smaller than
the energy effect and does not reverse it. Near the Coulomb barrier the capture cross
section depends exponentially on the barrier height, and a shift of a few hundred keV would
be observable in a barrier-distribution measurement. It must be stressed, however, that the
present calculation stops at $R=15.5$ fm, some $2.5$ fm outside the barrier, and that a
quantitative evaluation of the modified capture cross section is beyond its scope. What can
be stated is the scale of the effect at the separations at which the dinuclear system is
formed, not its value at the top of the barrier.

The situation is different for the dipole component of the response. It dominates the
polarization energy over the whole range of separations studied, and the DNS interaction
potential contains no corresponding degree of freedom, so that this contribution is absent
from the model rather than small within it. It is the direct analogue of the adiabatic
Coulomb polarization potential known from elastic scattering \cite{Baltz1979,Alder1975},
whose classical form $\Delta V=-\alpha_DZ_B^2e^2/2R^4$ reproduces the nearly constant
polarizability extracted in Sec. IV F. Its magnitude at the barrier cannot be established
here. The present calculation is restricted to $R\ge15.5$ fm, whereas the barrier of the
\Ca$+$\Pb~system is located near $R\simeq13$ fm, and both parts of the external field grow
rapidly between these two distances. A quantitative statement about the capture cross
section would require a two-centre treatment in the barrier region, together with the
removal of the centre-of-mass component of the dipole response discussed in Sec. IV F. The
present results should accordingly be regarded as establishing the scale of the effect
rather than as a quantitative modification of the entrance-channel potential. That scale, a
few hundred keV at $R\simeq15.5$ fm, is comparable to the spread produced by the different
radius parametrizations in use for the folding potential, and it is therefore in part
already absorbed in the empirical calibration of that potential.

The conclusion of this section is thus that for two spherical partners, and at the
separations at which the present calculation is applicable, the neglect of the density
polarization in the equations of relative motion is justified: the induced quadrupole
deformation lies below the precision to which the static deformations are known, and the
associated change of the radial force amounts to a few percent. It should be stressed that
this conclusion is specific to the case examined here. The induced deformation grows as the
separation decreases, and it is expected to be larger for statically deformed partners,
where the surfaces approach each other more closely in the tip orientation. Should the
induced deformation in those cases prove to be comparable with the accuracy of the
tabulated $\beta_2$ values, it would have to be included explicitly in the equations of
relative motion, either as an additional degree of freedom propagated alongside $R(t)$ or
through a polarization correction to the interaction potential. The verification of this
point for deformed nuclei and at separations closer to the barrier is left to future
work.

\section{Conclusions}

The polarization of the nucleon density of \Pb~and \Ca~in the dinuclear configuration has
been calculated microscopically by including the field of the partner nucleus in the
deformed Woods--Saxon mean field and solving the single-particle problem again. To our
knowledge this is the first calculation in which a Woods--Saxon single-particle problem is
rediagonalized with the combined Coulomb and DNS-consistent folded nuclear field of a
partner nucleus, and in which the induced density change is obtained explicitly and
decomposed into its Coulomb and nuclear parts. The double-folding and frozen Hartree-Fock
descriptions \cite{Satchler1979,Khoa2000,AdamianDNS1996} keep the densities frozen and do not
contain this correction; TDHF and DC-TDHF \cite{Umar2006,Washiyama2008,SimenelUmar2018}
contain it dynamically but yield the potential of the composite system rather than the
response of one partner, and do not separate the two components of the external field. The
external field consists of the Coulomb potential of the partner and of the nuclear
potential obtained by double folding of the Migdal density-dependent interaction, taken in
the same form as in the DNS approach, so that the polarization is consistent with the
potential which governs the relative motion.

The induced density change is localized in the surface region of both nuclei. It is
dominated by the repulsive Coulomb tidal field, which compresses the density along the
internuclear axis and produces a negative induced quadrupole moment. The nuclear part of
the external field acts in the opposite direction, but because the Migdal force has zero
range this contribution becomes appreciable only when the surfaces of the two nuclei
approach each other. As a result the modulus of $\Delta Q_{20}$ passes through a maximum
at $R\simeq17$ fm for \Pb~and at $R\simeq16$--$17$ fm for \Ca, and no change of sign
occurs within the range of separations accessible to the one-centre description. An
extrapolation of the exponential growth of the nuclear contribution places that change of
sign near $R\simeq15$ fm, so that a prolate polarization is to be expected in the barrier
region itself. The
Coulomb part of the response follows the $R^{-3}$ tidal law over the whole range. The
polarization energy amounts to about $-1.5$ MeV for the system at $R=15.5$ fm.
Expressed through the deformation parameter, the induced quadrupole deformations are of
the order of $10^{-3}$. The lighter partner is the more strongly polarized of the two in
relative terms, $\Delta\beta_2$ of \Ca~being about twice that of \Pb~at every separation,
although its induced quadrupole moment is eight times smaller in absolute value.

The comparison with the first-order treatment of the same external field shows that the
response of \Pb~is linear to within $10\%$ down to $R=17$ fm, whereas the response of the
lighter partner \Ca~saturates already at large separations because of the much stronger
field of \Pb. Both nuclei recover the linear-response result when the strength of the
external field is reduced, which confirms the internal consistency of the two treatments.

A quantitative criterion for the validity of the one-centre approximation has also been
established, which does not appear to have been formulated previously for this class of
calculation. The description remains valid as long as the external field is unable to
bind a nucleon at the surface of the nucleus under study, which for the \Ca$+$\Pb~system
corresponds to $R>15.3$ fm. Below this distance the barrier between the two single-particle
wells disappears, nucleons are transferred to the partner, and the result does not converge
with the size of the basis. Since the touching configuration of this system lies at
$R\simeq12.1$ fm, the region of the interaction potential which is most important for the
capture process is not accessible to a one-centre calculation, and a two-centre basis is
required. In the DNS approach the fragment densities are kept frozen and the polarization is not
included in the equations of relative motion. The present calculation provides the check
of that assumption which has so far been missing. For two spherical partners, and at the
separations at which the one-centre description is applicable, the neglect is justified,
since the induced quadrupole deformation is smaller than the uncertainty of the tabulated
static deformations and the associated change of the radial force is of the order of a few
percent. This result is specific to the case studied here. If for deformed partners, or at
separations closer to the barrier, the induced deformation is found to be comparable with
that uncertainty, it should be taken into account explicitly in the equations of relative
motion in future work.

The present study has been restricted to two spherical partners. This choice isolates the
induced polarization from the static deformation of the ground state and makes it possible
to establish the method and the conditions of its applicability. The natural continuation
is the treatment of statically deformed nuclei, for which two new effects are expected.
First, the external field acting on a deformed partner depends on the orientation of its
symmetry axis relative to the internuclear axis. For a prolate partner the surfaces come
closer at a given $R$ in the tip orientation, and since the nuclear part of the external
field varies over the short length scale of the density diffuseness, it is the nuclear
contribution rather than the Coulomb one that is strongly orientation dependent. The
polarization energy is therefore expected to be larger in the tip orientation, whereas the
induced quadrupole moment may well be smaller there, because the prolate nuclear
contribution then cancels more of the oblate Coulomb one. The separation at which the sign
of $\Delta Q_{20}$ changes should accordingly move outwards in the tip orientation. Second, the induced deformation adds to the static one and may either
increase or reduce the effective deformation entering the driving potential, depending on
the sign of the ground-state $\beta_2$. Since the deformation parameters of the fragments
and their orientations are among the main quantities determining the capture and
quasifission probabilities in the DNS approach \cite{Nasirov2005}, the polarization
corrections to them are of direct interest, and their calculation is planned as the next
step of this work.

\section{Acknowledgments}
The author is grateful to A. K. Nasirov for valuable discussions.

\bibliography{references}

@article{Moller1995,
title = {Nuclear Ground-State Masses and Deformations},
author = {P. Moller and J.R. Nix and W.D. Myers and W.J. Swiatecki},
journal = {At. Data Nucl. Data Tables},
volume = {59},
number = {2},
pages = {185-381},
year = {1995},
issn = {0092-640X},
doi = {https://doi.org/10.1006/adnd.1995.1002},
url = {https://www.sciencedirect.com/science/article/pii/S0092640X85710029}
}

@article{Adamian1997,
  title = {Friction coefficient for deep-inelastic heavy-ion collisions},
  author = {Adamian, G. G. and Jolos, R. V. and Nasirov, A. K. and Muminov, A. I.},
  journal = {Phys. Rev. C},
  volume = {56},
  issue = {1},
  pages = {373--380},
  numpages = {0},
  year = {1997},
  month = {Jul},
  publisher = {American Physical Society},
  doi = {10.1103/PhysRevC.56.373},
  url = {https://link.aps.org/doi/10.1103/PhysRevC.56.373}
}

@book{Migdal1983,
author = {A. B. Migdal},
title = {Theory of Finite Fermi Systems and Applications to Atomic Nuclei},
publisher = {Nauka},
address = {Moscow},
edition = {2nd},
year = {1983}
}

@article{Hofman2000,
  title = {The discovery of the heaviest elements},
  author = {Hofmann, S. and M\"unzenberg, G.},
  journal = {Rev. Mod. Phys.},
  volume = {72},
  issue = {3},
  pages = {733--767},
  numpages = {45},
  year = {2000},
  month = {Jul},
  publisher = {American Physical Society},
  doi = {10.1103/RevModPhys.72.733},
  url = {https://link.aps.org/doi/10.1103/RevModPhys.72.733}
}

@article{Nasirov2005,
title = {The role of orientation of nucleus symmetry axis in fusion dynamics},
author = {Avazbek Nasirov and Akira Fukushima and Yuka Toyoshima and Yoshihiro Aritomo and Akhtam
Muminov and Shuhrat Kalandarov and Ravshanbek Utamuratov},
journal = {Nucl. Phys. A},
volume = {759},
number = {3},
pages = {342-369},
year = {2005},
issn = {0375-9474},
doi = {https://doi.org/10.1016/j.nuclphysa.2005.05.152},
url = {https://www.sciencedirect.com/science/article/pii/S0375947405008572}
}

@article{Adamian2000,
title = {Isotopic dependence of fusion cross sections in reactions with heavy nuclei},
author = {Adamian, G. G. and Antonenko, N. V. and Scheid, W.},
journal = {Nuclear Physics A},
volume = {678},
number = {1},
pages = {24-38},
year = {2000},
issn = {0375-9474},
doi = {https://doi.org/10.1016/S0375-9474(00)00317-1},
url = {https://www.sciencedirect.com/science/article/pii/S0375947400003171}
}

@article{Antonenko1995,
  title = {Compound nucleus formation in reactions between massive nuclei: Fusion barrier},
  author = {Antonenko, N. V. and Cherepanov, E. A. and Nasirov, A. K. and Permjakov, V. P. and Volkov,
  V. V.},
  journal = {Phys. Rev. C},
  volume = {51},
  issue = {5},
  pages = {2635--2645},
  numpages = {0},
  year = {1995},
  month = {May},
  publisher = {American Physical Society},
  doi = {10.1103/PhysRevC.51.2635},
  url = {https://link.aps.org/doi/10.1103/PhysRevC.51.2635}
}

@article{Raman2001,
title = {TRANSITION PROBABILITY FROM THE GROUND TO THE FIRST-EXCITED 2+ STATE OF EVEN–EVEN NUCLIDES},
journal = {Atomic Data and Nuclear Data Tables},
volume = {78},
number = {1},
pages = {1-128},
year = {2001},
issn = {0092-640X},
doi = {https://doi.org/10.1006/adnd.2001.0858},
url = {https://www.sciencedirect.com/science/article/pii/S0092640X01908587},
author = {S. Raman and C. W. Nestor and P. Tikkanen}
}

@article{Giuliani2019,
  title = {Colloquium: Superheavy elements: Oganesson and beyond},
  author = {Giuliani, S. A. and Matheson, Z. and Nazarewicz, W. and Olsen, E. and Reinhard, P.-G. and
  Sadhukhan, J. and Schuetrumpf, B. and Schunck, N. and Schwerdtfeger, P.},
  journal = {Rev. Mod. Phys.},
  volume = {91},
  issue = {1},
  pages = {011001},
  numpages = {25},
  year = {2019},
  month = {Jan},
  publisher = {American Physical Society},
  doi = {10.1103/RevModPhys.91.011001},
  url = {https://link.aps.org/doi/10.1103/RevModPhys.91.011001}
}

@article{Nasirov2024,
  title = {Optimal colliding energy for the synthesis of a superheavy element with $Z=119$},
  author = {Nasirov, Avazbek and Kayumov, Bakhodir},
  journal = {Phys. Rev. C},
  volume = {109},
  issue = {2},
  pages = {024613},
  numpages = {10},
  year = {2024},
  month = {Feb},
  publisher = {American Physical Society},
  doi = {10.1103/PhysRevC.109.024613},
  url = {https://link.aps.org/doi/10.1103/PhysRevC.109.024613}
}

@article{Kayumov2022,
    author = "Kayumov, B. M. and Ganiev, O. K. and Nasirov, A. K. and Yuldasheva, G. A.",
    title = "{Analysis of the fusion mechanism in the synthesis of superheavy element 119 via the
    Cr54+Am243 reaction}",
    eprint = "2111.02861",
    archivePrefix = "arXiv",
    primaryClass = "nucl-th",
    doi = "10.1103/PhysRevC.105.014618",
    journal = "Phys. Rev. C",
    volume = "105",
    number = "1",
    pages = "014618",
    year = "2022"
}

@article{Cwiok1987,
 title = {Single-particle energies, wave functions, quadrupole moments and g-factors in an axially deformed woods-saxon potential with applications to the two-centre-type nuclear problems},
journal = {Computer Physics Communications},
volume = {46},
number = {3},
pages = {379-399},
year = {1987},
issn = {0010-4655},
doi = {https://doi.org/10.1016/0010-4655(87)90093-2},
url = {https://www.sciencedirect.com/science/article/pii/0010465587900932},
author = {S. Cwiok and J. Dudek and W. Nazarewicz and J. Skalski and T. Werner}
}

@article{Satchler1979,
   author = "Satchler, G. R. and Love, W. G.",
    title = "{Folding model potentials from realistic interactions for heavy-ion scattering}",
    doi = "10.1016/0370-1573(79)90081-4",
    journal = "Phys. Rept.",
    volume = "55",
    pages = "183--254",
    year = "1979"
}

@article{Khoa2000,
    author = "Khoa, Dao T. and Satchler, G. R.",
    title = "{Generalized folding model for elastic and inelastic nucleus{\textendash}nucleus scattering using realistic density dependent nucleon{\textendash}nucleon interaction}",
    doi = "10.1016/S0375-9474(99)00680-6",
    journal = "Nucl. Phys. A",
    volume = "668",
    pages = "3--41",
    year = "2000"
}

@article{Umar2006,
  title = {Heavy-ion interaction potential deduced from density-constrained time-dependent Hartree-Fock calculation},
  author = {Umar, A. S. and Oberacker, V. E.},
  journal = {Phys. Rev. C},
  volume = {74},
  issue = {2},
  pages = {021601(R)},
  numpages = {3},
  year = {2006},
  month = {Aug},
  publisher = {American Physical Society},
  doi = {10.1103/PhysRevC.74.021601},
  url = {https://link.aps.org/doi/10.1103/PhysRevC.74.021601}
}

@article{Washiyama2008,
  title = {Energy dependence of the nucleus-nucleus potential close to the Coulomb barrier},
  author = {Washiyama, Kouhei and Lacroix, Denis},
  journal = {Phys. Rev. C},
  volume = {78},
  issue = {2},
  pages = {024610},
  numpages = {10},
  year = {2008},
  month = {Aug},
  publisher = {American Physical Society},
  doi = {10.1103/PhysRevC.78.024610},
  url = {https://link.aps.org/doi/10.1103/PhysRevC.78.024610}
}

@article{SimenelUmar2018,
title = {Heavy-ion collisions and fission dynamics with the time-dependent Hartree–Fock theory and its extensions},
journal = {Progress in Particle and Nuclear Physics},
volume = {103},
pages = {19-66},
year = {2018},
issn = {0146-6410},
doi = {https://doi.org/10.1016/j.ppnp.2018.07.002},
url = {https://www.sciencedirect.com/science/article/pii/S0146641018300693},
author = {C. Simenel and A.S. Umar}
}

@article{Simenel2012,
  title     = "Nuclear quantum many-body dynamics",
  author    = "Simenel, C{\'e}dric",
  journal   = "Eur. Phys. J. A",
  publisher = "Springer Science and Business Media LLC",
  volume    =  48,
  number    =  11,
  pages     =  152,
  month     =  nov,
  year      =  2012,
  copyright = "https://www.springernature.com/gp/researchers/text-and-data-mining",
  language  = "en"
}

@article{Denisov2002,
  title     = "Entrance channel potentials in the synthesis of the heaviest
               nuclei",
  author    = "Denisov, V Yu and N{\"o}renberg, W",
  journal   = "Eur. Phys. J. A",
  publisher = "Springer Science and Business Media LLC",
  volume    =  15,
  number    =  3,
  pages     = "375--388",
  month     =  nov,
  year      =  2002,
  language  = "en"
}

@book{RingSchuck,
  title = {The Nuclear Many-Body Problem},
  author    = "Ring, Peter and Schuck, Peter",
  publisher = "Springer Berlin Heidelberg",
  address = {Berlin}, year = {1980}
}

@article{Angeli2013,
title = {Table of experimental nuclear ground state charge radii: An update},
journal = {Atomic Data and Nuclear Data Tables},
volume = {99},
number = {1},
pages = {69-95},
year = {2013},
issn = {0092-640X},
doi = {https://doi.org/10.1016/j.adt.2011.12.006},
url = {https://www.sciencedirect.com/science/article/pii/S0092640X12000265},
author = {I. Angeli and K.P. Marinova}
}

@article{deVries1987,
 title = {Nuclear charge-density-distribution parameters from elastic electron scattering},
journal = {Atomic Data and Nuclear Data Tables},
volume = {36},
number = {3},
pages = {495-536},
year = {1987},
issn = {0092-640X},
doi = {https://doi.org/10.1016/0092-640X(87)90013-1},
url = {https://www.sciencedirect.com/science/article/pii/0092640X87900131},
author = {H. {De Vries} and C.W. {De Jager} and C. {De Vries}}
}

@article{Wang2021,
 doi = {10.1088/1674-1137/abddaf},
url = {https://doi.org/10.1088/1674-1137/abddaf},
year = {2021},
month = {mar},
publisher = {Chinese Physical Society and the Institute of High Energy Physics of the Chinese Academy of Sciences and the Institute of Modern Physics of the Chinese Academy of Sciences and IOP Publishing Ltd},
volume = {45},
number = {3},
pages = {030003},
author = {Wang, Meng and Huang, W.J. and Kondev, F.G. and Audi, G. and Naimi, S.},
title = {The AME 2020 atomic mass evaluation (II). Tables, graphs and references*},
journal = {Chinese Physics C}
}

@article{Randrup1984,
title = {Dissipative resistance against changes in the mass asymmetry degree of freedom in nuclear dynamics: The completed wall-and-window formula},
journal = {Nuclear Physics A},
volume = {429},
number = {1},
pages = {105-115},
year = {1984},
issn = {0375-9474},
doi = {https://doi.org/10.1016/0375-9474(84)90151-9},
url = {https://www.sciencedirect.com/science/article/pii/0375947484901519},
author = {J. Randrup and W.J. Swiatecki}
}

@article{Baltz1979,
title = {Long range absorption and other optical-model effects from strong inelastic coupling},
journal = {Nuclear Physics A},
volume = {327},
number = {1},
pages = {221-249},
year = {1979},
issn = {0375-9474},
doi = {https://doi.org/10.1016/0375-9474(79)90326-9},
url = {https://www.sciencedirect.com/science/article/pii/0375947479903269},
author = {A.J. Baltz and N.K. Glendenning and S.K. Kauffmann and K. Pruess}
}

@book{Alder1975,
  title = {Electromagnetic Excitation},
  author = {Alder, K. and Winther, A.}, publisher = {North-Holland}, address = {Amsterdam}, year = {1975}}

@article{Tamii2011,
  title = {Complete Electric Dipole Response and the Neutron Skin in $^{208}\mathrm{Pb}$},
  author = {Tamii, A. and Poltoratska, I. and von Neumann-Cosel, P. and Fujita, Y. and Adachi, T. and Bertulani, C. A. and Carter, J. and Dozono, M. and Fujita, H. and Fujita, K. and Hatanaka, K. and Ishikawa, D. and Itoh, M. and Kawabata, T. and Kalmykov, Y. and Krumbholz, A. M. and Litvinova, E. and Matsubara, H. and Nakanishi, K. and Neveling, R. and Okamura, H. and Ong, H. J. and \"Ozel-Tashenov, B. and Ponomarev, V. Yu. and Richter, A. and Rubio, B. and Sakaguchi, H. and Sakemi, Y. and Sasamoto, Y. and Shimbara, Y. and Shimizu, Y. and Smit, F. D. and Suzuki, T. and Tameshige, Y. and Wambach, J. and Yamada, R. and Yosoi, M. and Zenihiro, J.},
  journal = {Phys. Rev. Lett.},
  volume = {107},
  issue = {6},
  pages = {062502},
  numpages = {5},
  year = {2011},
  month = {Aug},
  publisher = {American Physical Society},
  doi = {10.1103/PhysRevLett.107.062502},
  url = {https://link.aps.org/doi/10.1103/PhysRevLett.107.062502}
}

@article{Fearick2023,
  title = {Electric dipole polarizability of $^{40}\mathrm{Ca}$},
  author = {Fearick, R. W. and von Neumann-Cosel, P. and Bacca, S. and Birkhan, J. and Bonaiti, F. and Brandherm, I. and Hagen, G. and Matsubara, H. and Nazarewicz, W. and Pietralla, N. and Ponomarev, V. Yu. and Reinhard, P.-G. and Roca-Maza, X. and Richter, A. and Schwenk, A. and Simonis, J. and Tamii, A.},
  journal = {Phys. Rev. Res.},
  volume = {5},
  issue = {2},
  pages = {L022044},
  numpages = {6},
  year = {2023},
  month = {May},
  publisher = {American Physical Society},
  doi = {10.1103/PhysRevResearch.5.L022044},
  url = {https://link.aps.org/doi/10.1103/PhysRevResearch.5.L022044}
}

@article{AdamianDNS1996,
    author = "Adamian, G. G. and Antonenko, N. V. and Jolos, R. V. and Ivanova, S. P. and Melnikova, O. I.",
    title = "{Effective nucleus-nucleus potential for calculation of potential energy of a dinuclear system}",
    doi = "10.1142/S0218301396000098",
    journal = "Int. J. Mod. Phys. E",
    volume = "5",
    pages = "191--216",
    year = "1996"
}

@article{Kayumov2025,
  title = {Investigation of the formation of superheavy elements with atomic numbers 116 and 120 through $^{50}\mathrm{Ti}$-induced reactions},
  author = {Kayumov, Bakhodir},
  journal = {Phys. Rev. C},
  volume = {111},
  issue = {5},
  pages = {054619},
  numpages = {10},
  year = {2025},
  month = {May},
  publisher = {American Physical Society},
  doi = {10.1103/PhysRevC.111.054619},
  url = {https://link.aps.org/doi/10.1103/PhysRevC.111.054619}
}

\end{document}